\documentclass[12pt]{article}
\usepackage[totalheight = 23cm, totalwidth = 17cm]{geometry}
\usepackage{amssymb,amsmath,amsfonts,amsbsy,graphicx,bm,rotating}

\def\mathbi#1{\textbf{\em #1}}

\newcommand{\calH}{\mathcal{H}}
\newcommand{\calN}{\mathcal{N}}
\newcommand{\calO}{\mathcal{O}}

\begin{document}

\begin{titlepage}

\rightline{\footnotesize{APCTP-Pre2015-012}}
\vspace{-0.2cm}

\begin{center}

\vskip 1.0 cm

{\LARGE  \bf Dark energy and non-linear power spectrum}

\vskip 1.0cm

{\large
Sang Gyu Biern$^{a,b}$ and Jinn-Ouk Gong$^{b,c}$
}

\vskip 0.5cm

{\it
$^{a}$Department of Physics, Seoul National University, Seoul 151-747, Korea
\\
$^{b}$Asia Pacific Center for Theoretical Physics, Pohang 790-784, Korea
\\
$^{c}$Department of Physics, Postech, Pohang 790-784, Korea
}

\vskip 1.2cm

\end{center}

\begin{abstract}

We investigate the effects of homogeneous general dark energy on the non-linear matter perturbation in fully general relativistic context. The equation for the density contrast contains even at linear order new contributions which are non-zero for general dark energy. Taking into account the next-leading-order corrections, we derive the total power spectrum in real and redshift spaces. We find that the observable galaxy power spectrum deviates from the $\Lambda$CDM spectrum, which is nearly identical to that in the Einstein-de Sitter universe, and the relative difference is about $10\%$ on a scale of the baryon acoustic oscillations.

\end{abstract}

\end{titlepage}

\newpage

\setcounter{page}{1}

\section{Introduction}
\label{sec:intro}

After the discovery of the accelerated expansion of the universe from the observations on distant type Ia supernovae~\cite{SNIa}, among the biggest puzzles in cosmology is the identity of dark energy, the ``driving force'' of the acceleration that occupies about 68\% of the energy budget of the present universe~\cite{Ade:2015xua}. The simplest possibility is a small but non-vanishing cosmological constant~\cite{ccreview}, whose right amount would be then explained possibly by some unknown symmetry of quantum gravity or anthropic principle based on the landscape of vacua. Dynamical alternatives are non-trivial scalar fields such as Galileon~\cite{Nicolis:2008in}, or modification of gravity like massive gravity~\cite{deRham:2014zqa} or $f(R)$ theories~\cite{DeFelice:2010aj}.

While the identity of dark energy has a variety of theoretical origins, its properties are constrained by observations~\cite{Ade:2015xua}, from which we understand how dark energy has influenced the history of the universe: at very early times, for example when the cosmic microwave background (CMB) was generated, dark energy was completely negligible. At later stage when non-linearities in cosmic structure are developed, dark energy becomes significant and affects the evolution of gravitational instability. Therefore, the effects of dark energy should emerge more prominently at non-linear level. Possible signatures of dark energy in non-linear cosmic structure can be measured with more accurate data from large scale galaxy surveys into deeper redshift in near future such as HETDEX~\cite{HETDEX}, DESI~\cite{DESI}, LSST~\cite{LSST} and Euclid~\cite{Euclid}, which are to probe very large volume comparable to the horizon scale. This also demands fully general relativistic approaches beyond the Newtonian theory.

In this article, we study how homogeneous dark energy with a time-varying equation of state modifies non-linear matter power spectrum in the comoving gauge. The same topic was studied in the Newtonian perturbation theory but the effects of general dark energy are totally insignificant~\cite{Takahashi:2008yk}; see also~\cite{otherNL}. We find that in the presence of dynamical dark energy the next-to-leading corrections to the power spectrum can give rise to a deviation as large as a few percent from both the non-linear power spectrum in $\Lambda$CDM and the counterpart in the Einstein-de Sitter (EdS) universe at scales close to the baryon acoustic oscillations (BAO); meanwhile the power spectrum in $\Lambda$CDM is almost identical to that in the EdS universe with less than $0.1\%$ difference. This notable difference between the power spectrum in fully general relativistic context and the corresponding counterpart in the EdS universe, which is usually taken in the Newtonian theory, is thus a sharp feature of general dark energy models. For the power spectrum in the redshift space which is directly observable, the deviation becomes more prominent. At BAO scales the deviation from $\Lambda$CDM reaches about $10\%$ or so.

We also investigate how the homogenous dark energy affects the pure relativistic contributions coming from the curvature perturbation, which is heavily suppressed in the EdS universe~\cite{Jeong:2010ag,Biern:2014zja}. The curvature effects decrease more than $100\%$ from the EdS case in the presence of dark energy, including a cosmological constant, while their contribution is not significant at all.

\section{Modified equations with general dark energy}

We consider a flat Friedmann-Robertson-Walker universe as our background. The energy density of the universe consists of pressureless matter and non-interacting dark energy, which is a good approximation for the epoch of the evolution of cosmic structure after the generation of CMB until now. Furthermore, to simplify the analysis, we assume that dark energy is so homogeneously distributed that its perturbation is negligible. That is, dark energy contributes only to background; see however~\cite{Hwang:2009zj} for possible significance of dark energy perturbation. To fix the coordinate system, we choose the comoving gauge where $T^0{}_i = 0$ with the spatial metric $g_{ij} = a^2(\eta)(1+2\varphi)\delta_{ij}$. Here, $d\eta=dt/a$ is the conformal time. Note that the governing equations in the comoving gauge is known to coincide exactly with the Newtonian hydrodynamic equations up to second order in a zero pressure fluid~\cite{Noh:2004bc}. This is, however, not the case when we consider dark energy, because of the pressure associated with dark energy which plays a genuine relativistic role even at background level~\cite{Hwang:2007ni}. For example, the energy conservation equation has an additional contribution
\begin{equation}
-\kappa\bar\rho_m(1-\lambda) \quad \text{with} \quad \lambda \equiv (1+w)\left( 1-\frac{1}{\Omega_m} \right) \, .
\end{equation}
Here, $\kappa$ is the perturbation in the extrinsic curvature, an overbar means background, and $w$ is the equation of state of dark energy. $\lambda$ is thus an indicator of the effects of dark energy and is a key parameter in this work. Note that it vanishes in both EdS universe and $\Lambda$CDM.

The equation for the matter density contrast $\delta \equiv \delta\rho_m/\bar\rho_m$ is then obtained by combining the energy conservation and Raychaudhuri equations as
\begin{equation}
\label{eq:delta}
\delta'' + \left( \calH + \frac{\lambda'}{1-\lambda} \right)\delta' - \frac{3}{2}(1-\lambda)\calH^2\Omega_m\delta = \calN_N + \calN_\varphi + \calN_{\varphi'} + \calN_\lambda \, ,
\end{equation}
where a prime denotes a derivative with respect to the conformal time and $\calH \equiv a'/a$. Note that on the left hand side, where we have placed linear terms, we find new contributions with $\lambda$ that are absent in the conventional Newtonian perturbation theory. The terms on the right hand side are non-linear sources: $\calN_N$ denotes the non-linear Newtonian source~\cite{Takahashi:2008yk}, and the pure relativistic effect $\calN_\varphi$ originates from the curvature perturbation $\varphi$~\cite{Jeong:2010ag,Biern:2014zja}. The remaining two terms, $\calN_{\varphi'}$ and $\calN_\lambda$, represent respectively the evolution of $\varphi$ and the departure from $\Lambda$CDM and EdS, and are given by
\begin{align}
\label{eq:Nvarphi}
\calN_{\varphi'} & = -2\varphi'\delta^{,i}\beta,_i - 2\delta^{,i}\Delta^{-2} \left( \frac{7}{2}\Delta\varphi'\Delta\beta + 2\Delta\varphi_{,i}'\beta^{,i} - \frac{1}{2}\varphi_{,ij}'\beta^{,ij} + 2\varphi'\Delta^2\beta + 3\varphi_{,i}'\Delta\beta^{,i} \right)_{,i} \, ,
\\
\label{eq:Nlambda}
\calN_\lambda & = -\lambda \left[ \beta^{,ij}\beta_{,ij} - \frac{1}{3}(\Delta\beta)^2 \right] + \frac{\lambda}{1-\lambda} \left( \frac{4}{3}\delta'{}^2 + \delta''\delta + \calH\delta'\delta \right) + \frac{2\lambda'}{(1-\lambda)^2}\delta'\delta - \frac{\lambda'}{1-\lambda}\delta^{,i}\beta,_i
\nonumber\\
& \quad + \frac{\lambda}{1-\lambda} \left( \delta\delta^{,i}\beta_{,i}' + \frac{8}{3}\delta'\delta^{,i}\beta_{,i} + 2\delta\delta_{,i}'\beta^{,i} + \calH\delta\delta^{,i}\beta_{,i} \right) - \frac{\lambda(2-\lambda)}{(1-\lambda)^2} \left( \frac{8}{3}\delta'{}^2\delta + \delta''\delta^2 + \calH\delta'\delta^2 \right)\nonumber\\
& \quad - \lambda \left[ -\frac{2}{3}\Delta\beta\Delta\beta_r + 2\beta_{,ij}\beta_r^{,ij} - 4\varphi\beta^{,ij}\beta_{,ij} + \frac{4}{3}\varphi(\Delta\beta)^2 + \frac{4}{3}\varphi_{,i}\beta^{,i}\Delta\beta - 4\varphi_{,i}\beta_{,j}\beta^{,ij} \right]
\nonumber\\
& \quad + \frac{2\lambda'}{(1-\lambda)^2}\delta\delta^{,i}\beta_{,i} - \frac{3\lambda'}{(1-\lambda)^3}\delta'\delta^2 + \frac{\lambda'}{1-\lambda}\left( 2\varphi\delta^{,i}\beta_{,i} - \delta_{,i}\beta_r^{,i} \right)\, ,
\end{align}
where $\beta_r \equiv \Delta^{-2}\left( 7\Delta\beta\Delta\varphi/2 + 2\beta^{,i}\Delta\varphi_{,i} - \beta^{,ij}\varphi_{,ij}/2 + 2\varphi\Delta^2\beta + 3 \varphi^{,i}\Delta\beta_{,i} \right)$. Here, $\beta$ is the scalar perturbation in the $0i$ component of the metric given by $g_{0i} = a^2(\eta)\beta_{,i}$, $\Delta \equiv \delta^{ij}\partial_i\partial_j$ is the spatial Laplacian and $\Delta^{-1}$ is the inverse Laplacian operator. It is important to note that while the effects of the curvature perturbation appear only from third order~\cite{Hwang:2005he}, general dark energy modifies the equation from linear order as can be seen from \eqref{eq:delta} and \eqref{eq:Nlambda}. This is because dark energy changes the background expansion rate and thus its effects permeate throughout all order in perturbation.

At linear order, the solution of the curvature perturbation $\varphi$ can be found as
\begin{equation}
\varphi = -\Delta^{-1} \left( \frac{\calH}{1-\lambda}\delta' + \frac{3}{2}\calH^2\Omega_m\delta \right) = -\frac{\calH^2f}{1-\lambda} \left[ 1 + \frac{3}{2}(1-\lambda)\frac{\Omega_m}{f} \right] \Delta^{-1}\delta \, ,
\end{equation}
where $f \equiv d\log{D_1}/d\log{a}$ and $D_1$ is the linear growth function. In the EdS universe, the expansion rate of universe is same as the inhomogeneity growth rate at linear order: $D_1=a$ so that $\calH^2D_1$ is constant. Therefore, $\varphi$ is time-independent in the EdS universe. Similar conclusion of a constant $\varphi$ can be drawn in $\Lambda$CDM. On the other hands, $\varphi$ decreases when we consider general dark energy because of the faster expansion rate and slower growth of inhomogeneity. Thus, $\varphi'$ also indicates the effects of general dark energy.

\section{Non-linear power spectrum in real space}

To compute the power spectrum perturbatively, we expand the density and velocity divergence perturbations as $\delta = \delta_1 + \delta_2 + \delta_3 + \cdots$ and $\kappa=\kappa_1+\kappa_2+\kappa_3+\cdots$. The second and third order solutions are found in terms of the $n$-th order symmetric kernels $F_n$ along with time-dependent coefficients as
\begin{equation}
\label{eq:nl-delta_sol}
\begin{split}
\delta_2(\bm k,a) & = D_1^2(a)\sum_{i=a}^bc_{2i}(a) \int \frac{d^3q_1d^3q_2}{(2\pi)^3} \delta^{(3)}\left(\bm k - \bm q_{12}\right) F_{2i}(\bm q_1,\bm q_2) \delta_1(\bm q_1)\delta_1(\bm q_2) \, ,
\\
\delta_3(\bm k,a) & = D_1^3(a) \sum_{i=a}^f c_{3i}(a) \int \frac{d^3q_1d^3q_2d^3q_3}{(2\pi)^{3\cdot2}} \delta^{(3)}\left(\bm k - \bm q_{123}\right) F_{3i}(\bm q_1,\bm q_2,\bm q_3) \delta_1(\bm q_1)\delta_1(\bm q_2)\delta_1(\bm q_3)
\\
& \quad + D_1^3(a)\calH^2(a) \sum_{i=a}^b c_{3i}^\varphi(a) \int \frac{d^3q_1d^3q_2d^3q_3}{(2\pi)^{3\cdot2}} \delta^{(3)}\left(\bm k - \bm q_{123}\right) F_{3i}^\varphi(\bm q_1,\bm q_2,\bm q_3) \delta_1(\bm q_1)\delta_1(\bm q_2)\delta_1(\bm q_3) \, ,
\end{split}
\end{equation}
where $c_{ni} \equiv D_{ni}/D_1^n$, $c_{3i}^\varphi \equiv D_{3i}^\varphi/\left(D_1^3\calH^2\right)$ and $\bm q_{12\cdots n} \equiv \sum_1^n \bm q_i$. The superscript $\varphi$ means originated from $\varphi$. The solutions for $\kappa_n/\calH$ have exactly the same form as $\delta_n$, while the time-dependent coefficients are replaced by $k_{ni} \equiv K_{ni}/(\calH D_1^n)$ with $K_{ni}$ being $i$-th growth function of $\kappa_n$. In the EdS universe, the coefficients are fixed as certain numbers: $c_{2a}=3/7$, $k_{2a}=-1/7$, $c_{2b}=2/7$, $k_{2b}=4/7$, $c_{3a}=k_{3a}=c_{3b}=k_{3b}=1/2$, $c_{3c}=k_{3c}=2/7$, $c_{3d}=-2/63$, $k_{3d}=-1/21$, $c_{3e}=-1/9$, $k_{3e}=8/21$, $c_{3f}=8/63$, $k_{3f}=-2/21$, $c_{3a}^\varphi=k_{3a}^\varphi=-5/4$, $c_{3b}^\varphi=-5/14$, and $k_{3b}^\varphi=-20/7$. Since the universe is close to EdS when $\delta$ begins to grow, these numbers are set to be the initial conditions for these coefficients and we evolve them. Note that in $\delta_3$, unlike $\varphi$, we cannot separate the contributions originated from $\lambda$ because as stated in the previous section dark energy changes the background expansion rate. Also note that the contribution from $\varphi$ carries $\calH = aH$, which can be identified as the comoving horizon scale $k_H$.

The non-linear power spectrum is defined by $\left\langle \delta(\mathbi{k})\delta(\mathbi{k}') \right\rangle \equiv (2\pi)^3\delta^{(3)}(\mathbi{k}+\mathbi{k}')P(k)$. Under the perturbative expansion of $\delta$, we can write $P = P_{11} + P_{22} + P_{13} + \cdots$, where the subscript denotes the order of perturbations being correlated, so that $P_{11}$ is the linear power spectrum between two $\delta_1$'s and $P_{22}+P_{13}$ is the next-to-leading order, viz. one-loop power spectrum. The power spectrum with one-loop corrections is
\begin{align}
\label{eq:nlP}
P(k,a) & = D_1^2(a)P_{11}(k)
\nonumber\\
& \quad + D_1^4(a) \frac{k^3}{8\pi^2} \int_0^\infty dr P_{11}(kr) \int _{-1}^1dx \frac{\left\{ c_{2a} \left[ -x + r\left( -1 + 2x^2 \right) \right] + 2c_{2b} x\left( -1 + rx \right) \right\}^2}{\left( 1 + r^2 - 2rx \right)^2}
\nonumber\\
& \hspace{12.5em} \times P_{11}\left( k\sqrt{1+r^2-2rx} \right)
\nonumber\\
& \quad + D_1^4(a) \frac{k^3}{96\pi^2} P_{11}(k) \int_0^\infty dr \frac{P_{11}(kr)}{r^3} \bigg\{ 12r (c_{3d}+c_{3e}) - 4r^3 \left[ -6 c_{3a}+8 c_{3b}+6 c_{3c}+11 (c_{3d}+c_{3e}) \right]
\nonumber\\
& \qquad\qquad +12 r^7 (-2 c_{3a}+2 c_{3c}+c_{3d}+c_{3e}) - 4r^5 \left[ -24 c_{3a}+8 c_{3b}+16 c_{3c}+11 (c_{3d}+c_{3e}) \right]
\nonumber\\
& \qquad\qquad \left. +6 \left(r^2-1\right)^3 \left[ r^2 (-2 c_{3a}+2 c_{3c}+c_{3d}+c_{3e})-c_{3d}-c_{3e} \right] \log \left|\frac{1-r}{1+r}\right|
\right\}
\nonumber\\
& \quad + D_1^4(a)\calH^2(a) \frac{k}{16\pi^2} P_{11}(k) \int_0^\infty dr \frac{P_{11}(kr)}{r^3} \bigg\{ 8r \left[ 5 \left(r^2-2\right) r^2+9 \right] c_{3b}^\varphi+12r^3 \left(5-3 r^2\right) c_{3a}^\varphi
\nonumber\\
& \qquad\qquad \left. -2\left(r^2-1\right) \left[ r^4 \left( 9 c_{3a}^\varphi-10 c_{3b}^\varphi \right) + r^2 \left( 15 c_{3a}^\varphi-8 c_{3b}^\varphi \right)+18 c_{3b}^\varphi \right] \log \left|\frac{1-r}{1+r}\right|
\right\}
\nonumber\\
& \equiv P_{11} (k,a)+ P_{22}(k,a) + P_{13}(k,a) + P_{13}^\varphi(k,a) \, .
\end{align}
Here, $r$ and $x$ represent respectively the magnitude of the dummy integration momentum by $q \equiv rk$ with $0 \leq r \leq \infty$ and the cosine between $\mathbi{k}$ and $\mathbi{q}$ by $\mathbi{k}\cdot\mathbi{q} = kqx = k^2rx$ with $-1 \leq x \leq 1$. Note that in the EdS universe and $\Lambda$CDM, $P_{13}$ becomes purely Newtonian and $P_{13}^\varphi$ alone carries the relativistic effects~\cite{Jeong:2010ag}.

\begin{sidewaysfigure}[!]
\centering
 \includegraphics[width=0.48\textwidth]{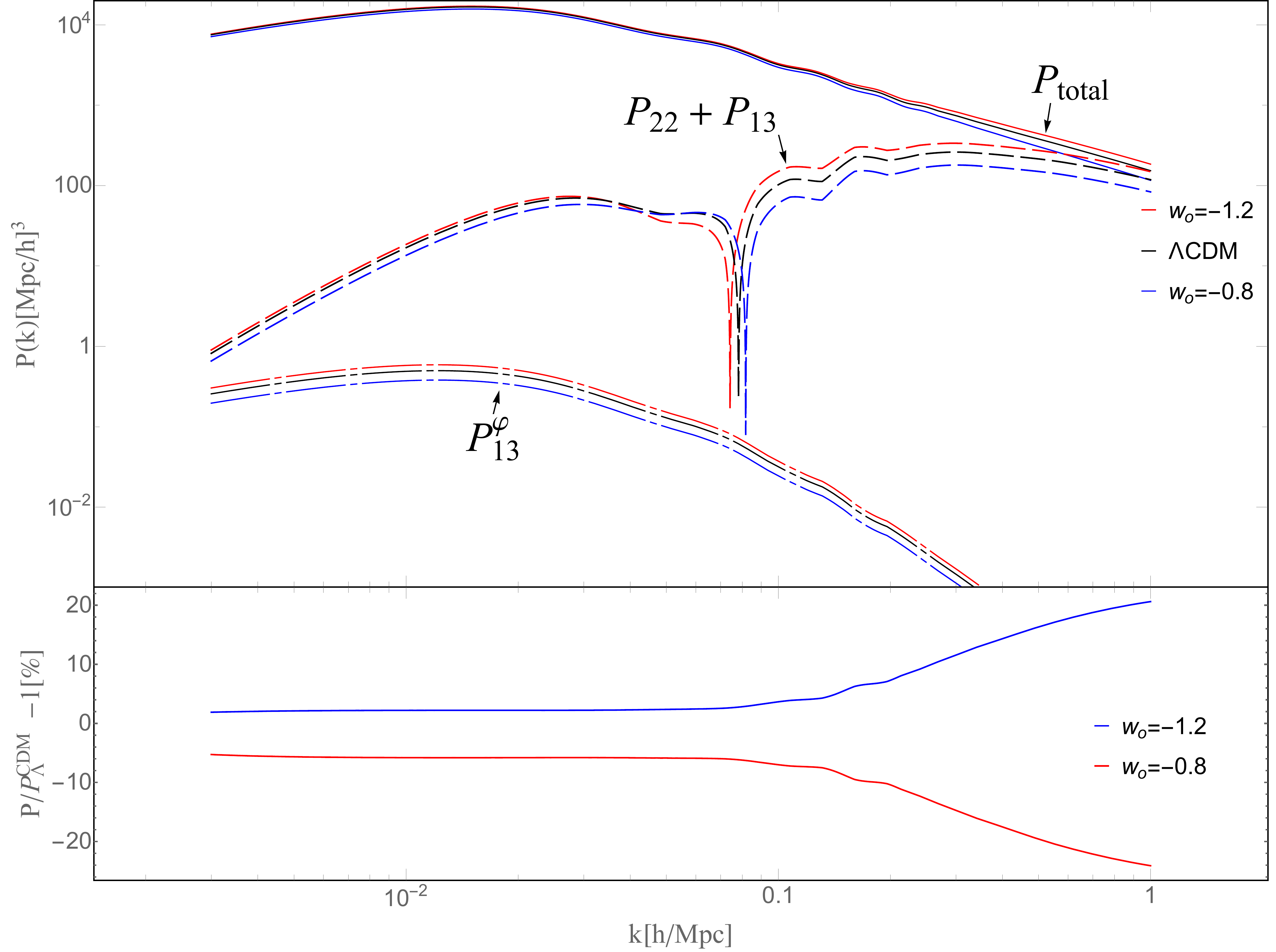}
 \hspace{0.5em}
 \includegraphics[width=0.48\textwidth]{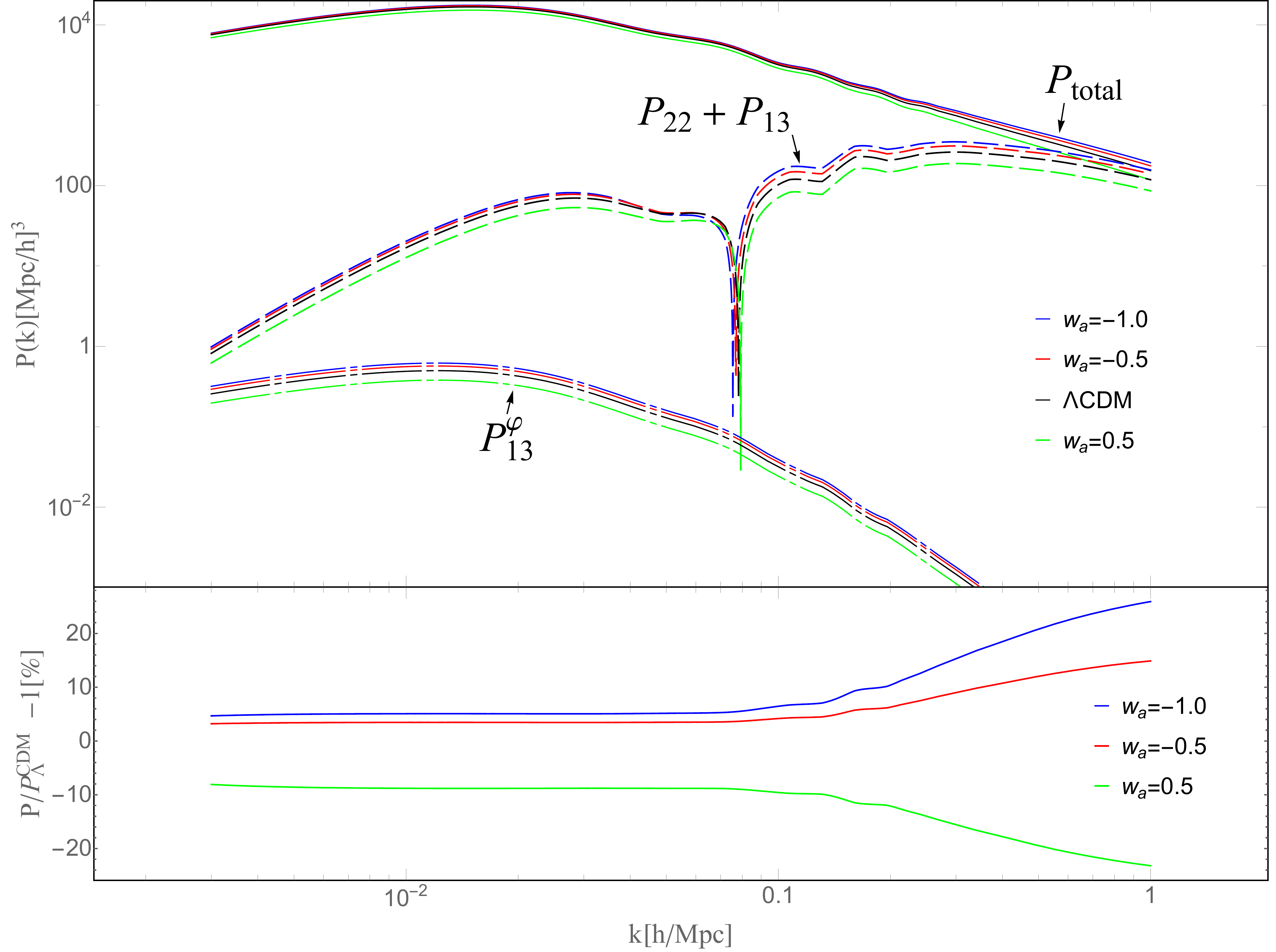}
 \caption{Total power spectra \eqref{eq:nlP} (solid lines) and the next-to-leading corrections $P_{22}+P_{13}$ (dashed lines) and $P_{13}^\varphi$ (dot-dashed lines) for various dark energy models (top panels) and the relative difference from $\Lambda$CDM (bottom panels). In the left panel we fix $w_a=0$ and vary $w_0$ as $w_0=-0.8$, $-1$ and $-1.2$, while in the right panel $w_0=-1$ is fixed and $w_a=0.5$, 0, $-0.5$ and $-1$ are presented.}
 \label{fig:LCDM}
\end{sidewaysfigure}

To evaluate the power spectrum \eqref{eq:nlP} for general dark energy, we parametrize the equation of state as~\cite{eos}
\begin{equation}
\label{eq:eos}
w(a) = w_0 + (1-a)w_a \, .
\end{equation}
Then we calculate the linear power spectrum $P_{11}$ from \texttt{CAMB}~\cite{Lewis:1999bs} with the best-fit parameters from the Planck 2015 data~\cite{Ade:2015xua}, which is used to evaluate the next-to-leading contributions $P_{22}+P_{13}+P_{13}^\varphi$ in \eqref{eq:nlP}. In Figure~\ref{fig:LCDM}, we present in the top panels the total power spectra for different dark energy models at $z=0$ along with the one-loop components $P_{22}+P_{13}$ and $P_{13}^\varphi$. In the left (right) column, with a fixed value of $w_a=0$ ($w_0=-1$) we vary $w_0$ ($w_a$): $w_0=-0.8$, $-1.0$ ($\Lambda$CDM) and $-1.2$ with $w_a=0$ in the left panel, and $w_a=0.5$, 0 ($\Lambda$CDM), $-0.5$ and $-1.0$ with $w_0=-1.0$ in the right panel respectively. In the bottom panels of Figure~\ref{fig:LCDM} the relative differences of the dark energy models from $\Lambda$CDM are presented. The deviation from $\Lambda$CDM is not significant and nearly constant on large scales ($k \lesssim 0.1h/$Mpc), but it increases from $k \simeq 0.1h$/Mpc and the relative difference becomes more or less 10\% at $k \sim 0.2h$/Mpc and even larger on smaller scales. We can note that the power spectrum for $w_0>-1$ ($w_0<-1$) in the left panel and $w_a>0$ ($w_a<0$) in the right panel is smaller (larger) than that in $\Lambda$CDM. Since we normalize $\Omega_{m0}$ as the Planck 2015 best-fit value, the energy density of dark energy increases (decreases) when $w<-1$ ($w>-1$). Thus the matter-dark energy equality is later for $w<-1$ than for $w>-1$, although the expansion rate for $w<-1$ is faster than that for $w>-1$ at present. As a result, the inhomogeneity can grow more for $w<-1$ than $w>-1$.

\begin{sidewaysfigure}[!]
\centering
 \includegraphics[width=0.48\textwidth]{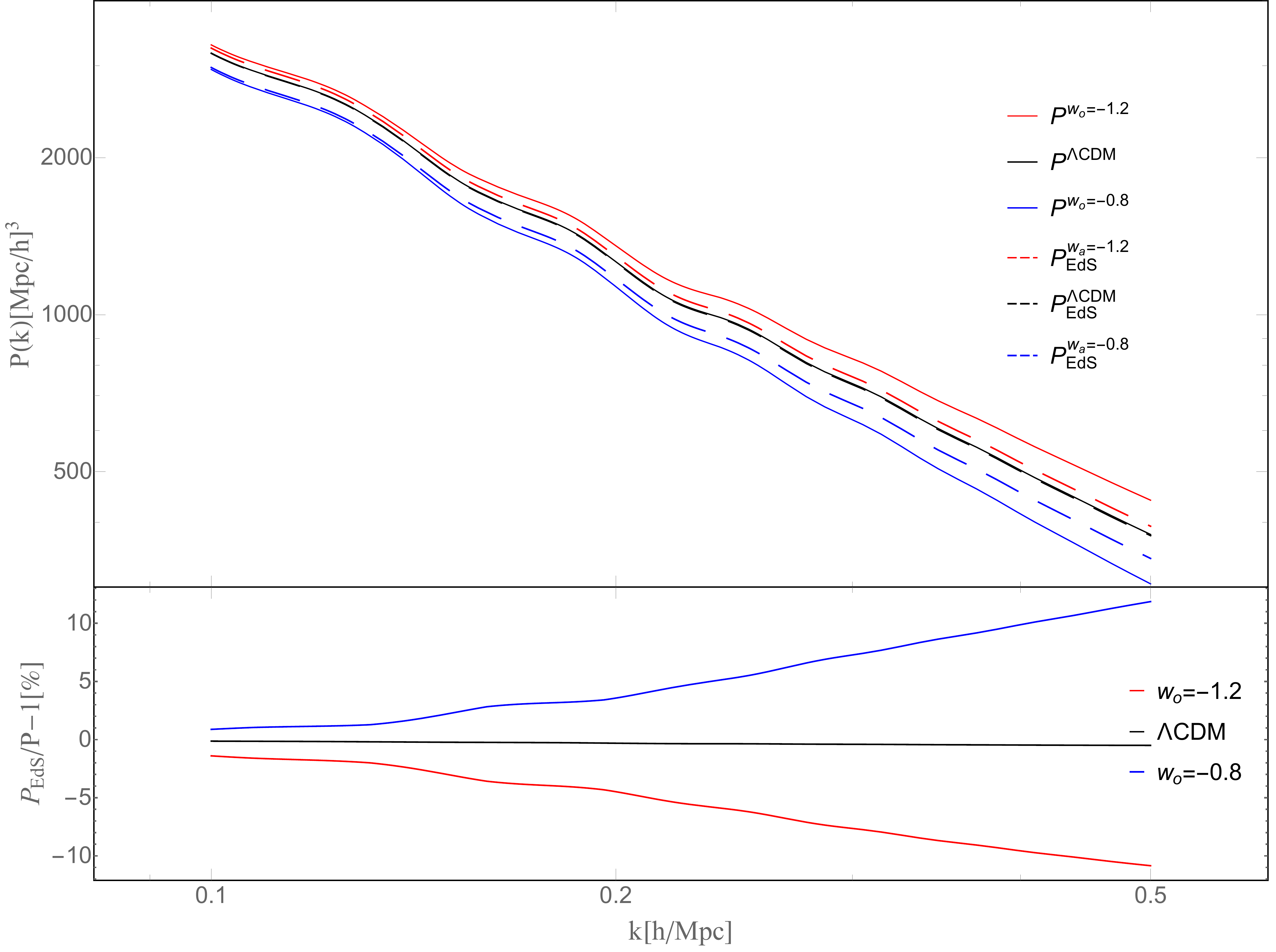}
 \hspace{1em}
 \includegraphics[width=0.48\textwidth]{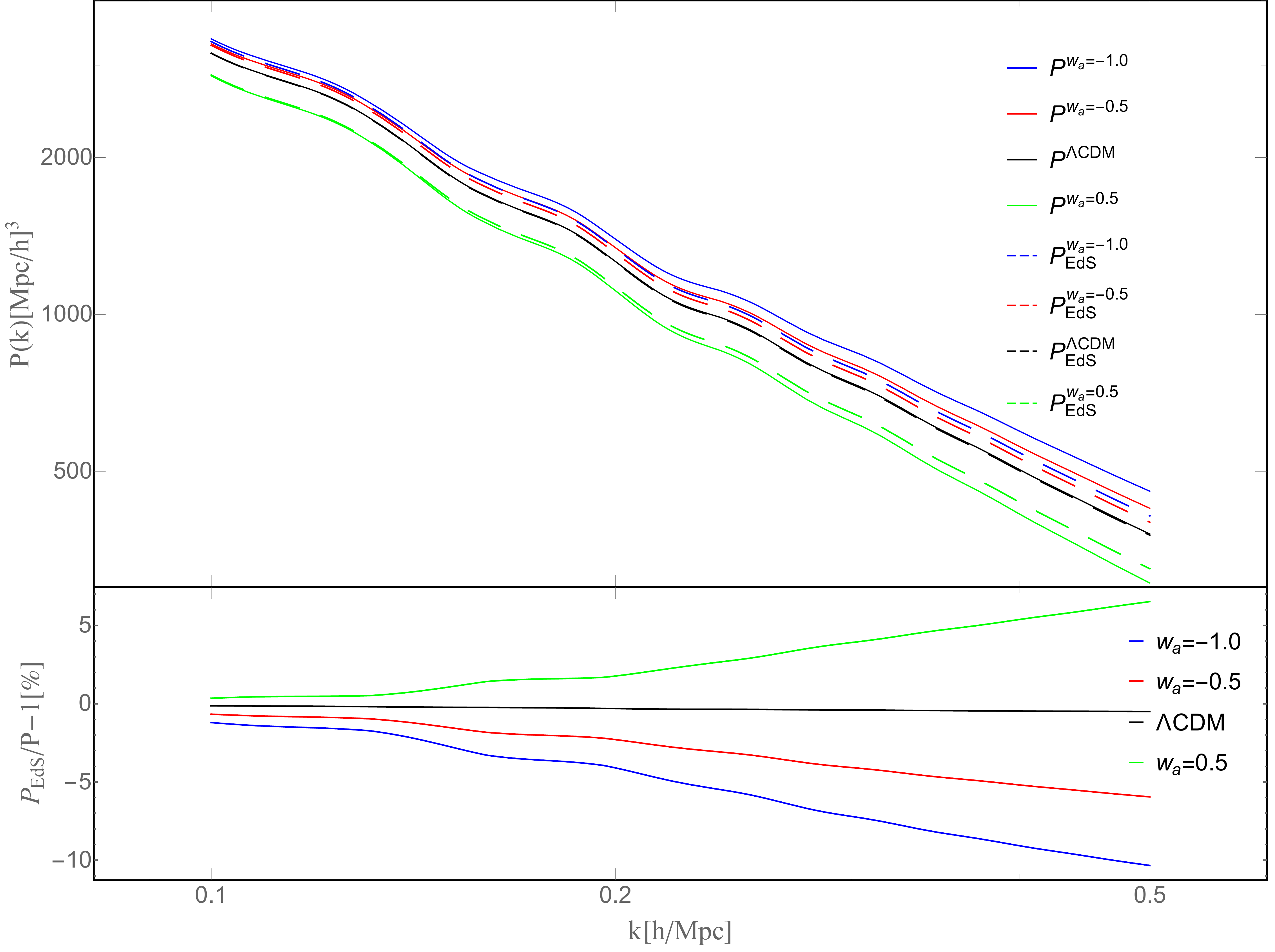}
 \caption{Total power spectra \eqref{eq:nlP} (solid lines) and the corresponding counter EdS spectra (dashed lines) for various dark energy models (top panels) and the relative difference (bottom panels). In the left panel we fix $w_a=0$ and vary $w_0$ as $w_0=-0.8$, $-1$ and $-1.2$, while in the right panel $w_0=-1$ is fixed and $w_a=0.5$, 0, $-0.5$ and $-1$ are presented.}
 \label{fig:EdS}
\end{sidewaysfigure}

Usually, one uses the non-linear power spectrum in the EdS universe to an arbitrary cosmological model by replacing the scale factor with the linear growth factor of that model, while leaving the time-independent part of the spectrum identical to the EdS one:
\begin{equation}
\label{eq:P_EdS}
P(k,a) = D_1^2(a)P_{11}(k) + D_1^4(a) \left[ P_{22}(k) + P_{13}(k) \right]_\text{EdS} \, .
\end{equation}
Compared with the {\em true} power spectrum \eqref{eq:nlP}, this is obviously inconsistent because in general dark energy models the non-linear kernels $F_i$'s as well as the time-dependent coefficients $c_i$'s of the non-linear solutions \eqref{eq:nl-delta_sol} are different from the EdS counterparts. In Figure~\ref{fig:EdS}, we compare the power spectra for different dark energy models with the corresponding counter EdS power spectra \eqref{eq:P_EdS}. As in Figure~\ref{fig:LCDM}, in the left (right) panel we fix $w_a=0$ ($w_0=-1$) and vary $w_0$ ($w_a$). The top panels show the total power spectra, while in the bottom panel the relative differences of the dark energy models from the corresponding EdS counterparts are presented. Since both $\calN_{\varphi'}$ and $\calN_\lambda$ vanish for $\Lambda$CDM, the relative difference of the total power spectrum in $\Lambda$CDM from the corresponding EdS one is only from $\calN_\varphi$ and is very small, $\calO\left(10^{-2}-10^{-1}\right)$\% over all scales~\cite{Jeong:2010ag}. On the other hand, the power spectrum in general dark energy differs substantially from the EdS counterpart because of the large $\calN_\lambda$ contribution: the relative difference notably increases from $k \approx 0.1h$/Mpc and becomes larger than 10\% for $k \geq 0.2h$/Mpc. While even larger deviations on smaller scales may not be reliable because of strong non-linear effects, the analytic non-linear power spectrum \eqref{eq:nlP} gives very good agreement with numerical simulations in weakly non-linear regime~\cite{Jeong:2006xd}.

It is also interesting to take a closer look at $P_{13}^\varphi$. In the top panels of Figure~\ref{fig:LCDM}, we separately show $P_{13}^\varphi$ for different dark energy models. While $P_{22}+P_{13}$ contributes more significantly on smaller scales $k \gtrsim 0.1h$/Mpc, $P_{13}^\varphi$ extends to larger scales. Furthermore, even for $\Lambda$CDM the relative difference from the counter EdS power spectrum is larger than $100\%$, i.e. always smaller in the presence of any form of dark energy on all scales. Thus, $P_{13}^\varphi$ is strongly influenced by dark energy and is in principle interesting, but its magnitude is heavily suppressed.

\section{Non-linear power spectrum in redshift space}

The power spectrum presented in the previous section is computed in the real space. In reality, however, we identify the position in terms of redshift. Thus to extract observational prospect, we need to consider the power spectrum in the redshift space which is directly related to observations.

The proper transformation from the real space to the redshift space includes all relativistic effects such as gravitational lensing and the Sachs-Wolfe effects. Note that although the real space solutions in the previous section are \emph{not} gauge-invariant, those in the redshift space obtained from this transformation are gauge-invariant~\cite{g-inv}. In this article, we take into account only the Doppler effect, because the other relativistic effects are suppressed at the BAO scales in which we are interested~\cite{g-inv,obsPg}. Then the density contrast in the redshift space is written as, up to third order~\cite{delta-redshift},
\begin{align}
\delta_s & = \delta _r-\partial_\|U+(\partial_\|U)^2-\delta_r\partial_\|U-\Delta U \partial_\|\delta_r+\Delta U \partial_\|^2U+\delta_r(\partial_\| U)^2
\nonumber\\
& \quad -(\partial_\|U)^3-3\Delta U\partial_\|U\partial_\|^2U+2\Delta U\partial_\|U\partial_\|\delta_r+\Delta U\delta_r\partial_\|^2U+\frac{1}{2}(\Delta U)^2\partial_\|^2\delta_r-\frac{1}{2}(\Delta U)^2\partial_\|^3 U \, ,
\end{align}
where $\delta_s$ ($\delta_r$) means the galaxy density contrast in the redshift (real) space. With a constant bias factor $b$, we can write $\delta_r = b\delta$~\cite{Kaiser:1984sw} with $\delta$ given by \eqref{eq:nl-delta_sol}. Since in the comoving gauge the coordinate time is equivalent to proper time, the bias factor may show up as it is without further manipulations~\cite{proper time}. The potential $U(\mathbi{r})$ is defined by $U \equiv \hat{\mathbi{n}}\cdot\mathbi{v}/\calH$, where $\mathbi{v}$ and $\hat{\mathbi{n}}$ are respectively the peculiar velocity and the unit vector along the line-of-sight direction (LoS). Since $\kappa = -\nabla\cdot\mathbi{v}$, $U$ can be written as $U = -\partial_\|\Delta^{-1}\kappa/\calH$, where $\partial_\| \equiv \hat{\mathbi{n}}\cdot\nabla$ is a partial differential operator along LoS.

Then, we can write the observable galaxy power spectrum in the redshift space as
\begin{equation}
\label{eq:Pg}
P_s(k,\mu,a)=P_{s11}(k,\mu,a)+P_{s22}(k,\mu,a)+P_{s13}(k,\mu,a) \, ,
\end{equation}
where $\mu \equiv \hat{\mathbi{n}}\cdot\mathbi{k}/k$ is the cosine between $\hat{\mathbi{n}}$ and $\mathbi{k}$. Note that \eqref{eq:Pg} is no longer isotropic even though the power spectrum in the real space is so. The power spectrum along LoS ($\mu=1$) is dominant, while the one with $\mu=0$, i.e. perpendicular to LoS is the same as \eqref{eq:nlP}. What is important is that hence the deviation from $\Lambda$CDM becomes more significant for the LoS power spectrum. In Table~\ref{table}, we give the relative differences of various dynamical dark energy model from $\Lambda$CDM. As can be seen, the deviation is even more enhanced compared to the real space power spectrum discussed in the previous section, as large as $10\%$ at around BAO scales, well within the expected observational accuracy for future surveys~\cite{prospect}.

\begin{table}[h]
\centering
\begin{tabular}{|c|c|c|c|}
\hline
\multicolumn{3}{|c|}{$w_a=0$ and varying $w_0$}\\ \hline\hline
 $k~[h$/Mpc]&$w_0=-1.2$  & $w_0=-0.8$ \\ \hline
 0.1               & 6.8$\%$      & -10.2$\%$  \\ \hline \cline{2-3}
0.2                & 11.6$\%$    & -15.0$\%$ \\ \hline  \cline{2-3}
0.3                &16.0$\%$     &-19.4$\%$  \\ \hline
\end{tabular}
\hspace{0.2cm}
\begin{tabular}{|c|c|c|c|}
\hline
\multicolumn{4}{|c|}{$w_0=-1$ and varying $w_a$}\\ \hline\hline
$k~[h$/Mpc]& $w_a=-1.0$ &$w_a=-0.5$ & $w_a=0.5$ \\ \hline \cline{2-3}
 0.1 & 9.5$\%$  & 5.8$\%$  &   -11.5$\%$ \\ \hline \cline{2-3}
0.2 &  14.9$\%$ & 8.8$\%$ & -15.3$\%$  \\ \hline  \cline{2-3}
0.3 &  20.1$\%$ &  11.6$\%$ &   -19.0$\%$ \\ \hline
\end{tabular}
\caption{Relative differences between various dark energy models and $\Lambda$CDM $P_s/P_{\Lambda\text{CDM}}-1[\%]$ with $\mu=1$, $z=0$ and $b=2$.}
\label{table}
\end{table}

\begin{sidewaysfigure}[!]
\centering
 \includegraphics[width=0.48\textwidth]{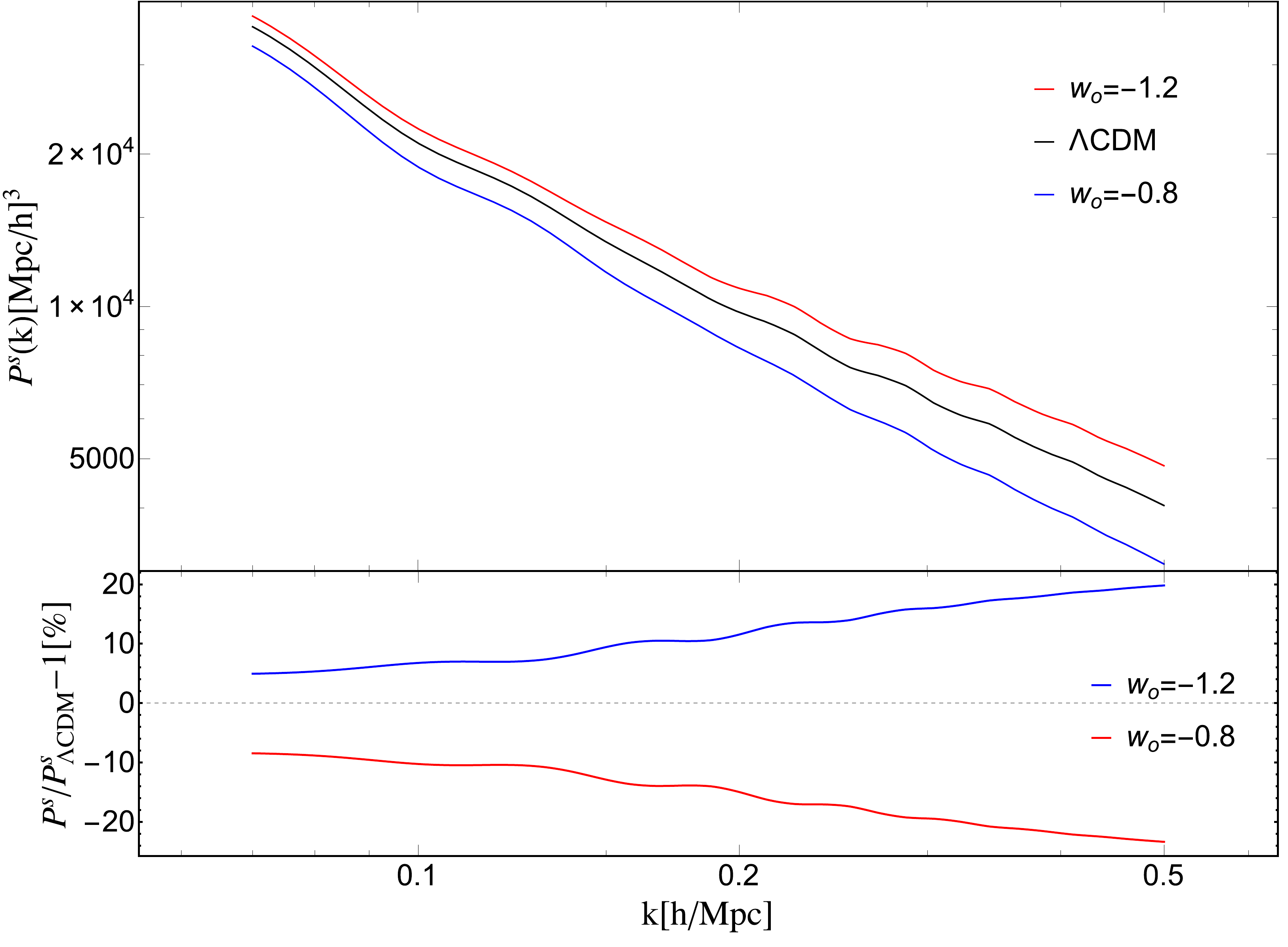}
 \hspace{1em}
 \includegraphics[width=0.48\textwidth]{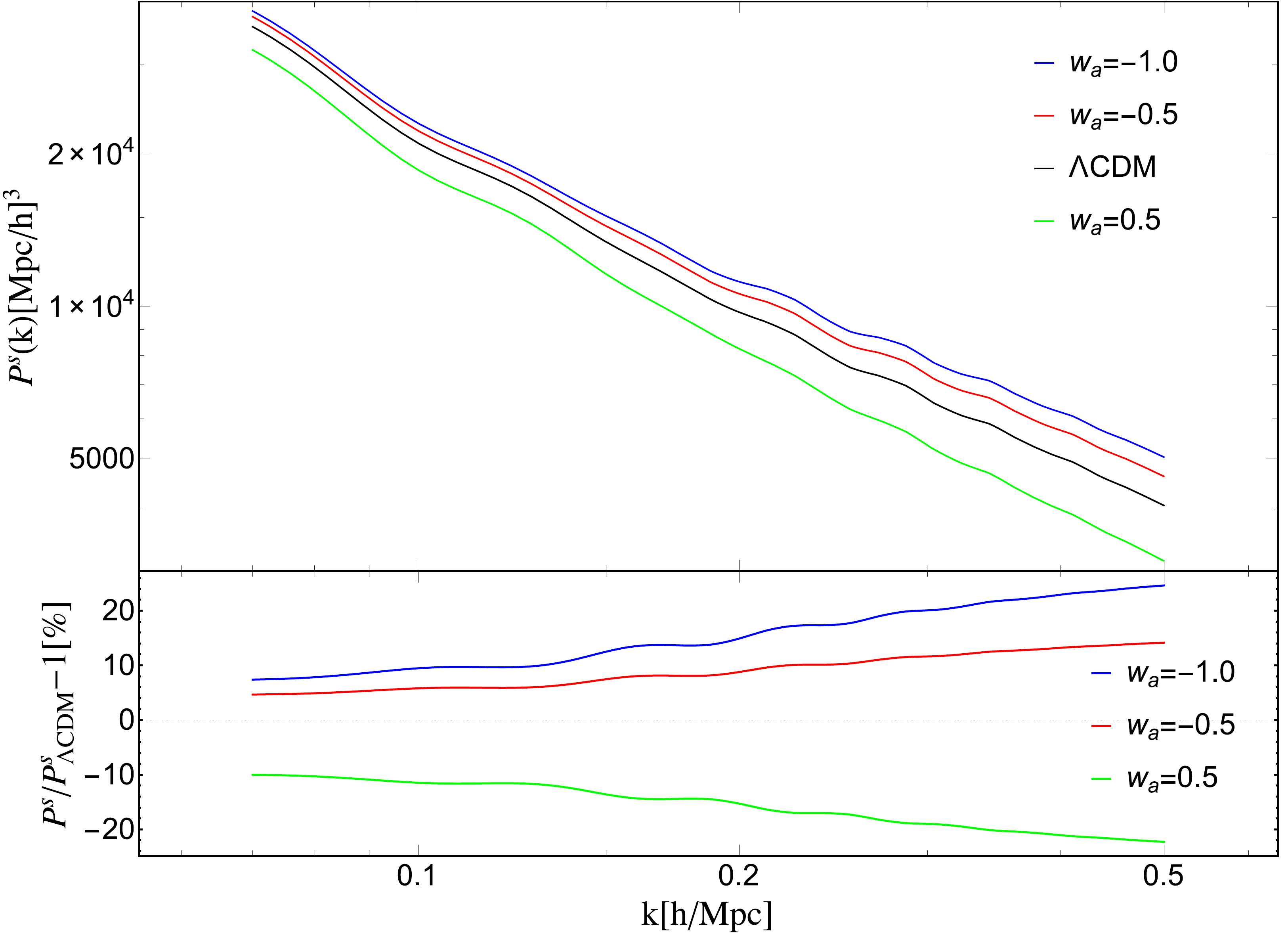}
 \caption{Observable power spectra \eqref{eq:Pg} with $\mu=1$, $z=0$ and $b=2$ for various dark energy models (top panels) and the relative difference (bottom panels). In the left panel we fix $w_a=0$ and vary $w_0$ as $w_0=-0.8$, $-1$ and $-1.2$, while in the right panel $w_0=-1$ is fixed and $w_a=0.5$, 0, $-0.5$ and $-1$ are presented.}
 \label{fig:RSD}
\end{sidewaysfigure}

\section{Conclusions}

In this article, we have studied the non-linear power spectrum in the presence of general, homogeneous dark energy in the context of fully general relativistic approach. We have derived the power spectra in the real and in the redshift space. We have found that in general dark energy different from a cosmological constant results in substantially large deviations from the conventional expectation in the observable power spectrum, as large as $10\%$, via the next-to-leading corrections on scales $k \gtrsim 0.1h$/Mpc.This is quite close to the BAO scales, thus our finding should be useful to constrain the properties of dark energy with the large scale BAO surveys.

\subsection*{Acknowledgements}

We thank Jai-chan Hwang, Donghui Jeong, Seokcheon Lee, Hyerim Noh, Soo-Jong Rey and Jaiyul Yoo for helpful discussions.
We are grateful to the Focus Research Program ``The origin and evolution of the Universe'', supported by the Asia Pacific Center for Theoretical Physics, while this work was under progress.
We acknowledge the Max-Planck-Gesellschaft, the Korea Ministry of Education, Science and Technology, Gyeongsangbuk-Do and Pohang City for the support of the Independent Junior Research Group at the Asia Pacific Center for Theoretical Physics. We are also supported by a Starting Grant through the Basic Science Research Program of the National Research Foundation of Korea (2013R1A1A1006701).


\begin{thebibliography}{99}




\bibitem{SNIa}
  A.~G.~Riess {\it et al.}  [Supernova Search Team Collaboration],
  Astron.\ J.\  {\bf 116}, 1009 (1998)
  [astro-ph/9805201]~;
  S.~Perlmutter {\it et al.}  [Supernova Cosmology Project Collaboration],
  Astrophys.\ J.\  {\bf 517}, 565 (1999)
  [astro-ph/9812133].


\bibitem{Ade:2015xua}
  P.~A.~R.~Ade {\it et al.}  [Planck Collaboration],
  arXiv:1502.01589 [astro-ph.CO].


\bibitem{ccreview}
  S.~Weinberg,
  Rev.\ Mod.\ Phys.\  {\bf 61}, 1 (1989)~;
  T.~Padmanabhan,
  Phys.\ Rept.\  {\bf 380}, 235 (2003)
  [hep-th/0212290].


\bibitem{Nicolis:2008in}
  A.~Nicolis, R.~Rattazzi and E.~Trincherini,
  Phys.\ Rev.\ D {\bf 79}, 064036 (2009)
  [arXiv:0811.2197 [hep-th]].


\bibitem{deRham:2014zqa}
  C.~de Rham,
  Living Rev.\ Rel.\  {\bf 17}, 7 (2014)
  [arXiv:1401.4173 [hep-th]].


\bibitem{DeFelice:2010aj}
  A.~De Felice and S.~Tsujikawa,
  Living Rev.\ Rel.\  {\bf 13}, 3 (2010)
  [arXiv:1002.4928 [gr-qc]].


\bibitem{HETDEX}
http://www.hetdex.org


\bibitem{DESI}
http://desi.lbl.gov


\bibitem{LSST}
http://www.lsst.org/lsst/


\bibitem{Euclid}
http://sci.esa.int/euclid/


\bibitem{Takahashi:2008yk}
  R.~Takahashi,
  Prog.\ Theor.\ Phys.\  {\bf 120}, 549 (2008)
  [arXiv:0806.1437 [astro-ph]].


\bibitem{otherNL}
  S.~Anselmi, G.~Ballesteros and M.~Pietroni,
  JCAP {\bf 1111}, 014 (2011)
  [arXiv:1106.0834 [astro-ph.CO]]~;
  S.~Lee, C.~Park and S.~G.~Biern,
  Phys.\ Lett.\ B {\bf 736}, 403 (2014)
  [arXiv:1407.7325 [astro-ph.CO]].


\bibitem{Jeong:2010ag}
  D.~Jeong, J.~O.~Gong, H.~Noh and J.~c.~Hwang,
  Astrophys.\ J.\  {\bf 727}, 22 (2011)
  [arXiv:1010.3489 [astro-ph.CO]]~;


\bibitem{Biern:2014zja}
  S.~G.~Biern, J.~O.~Gong and D.~Jeong,
  Phys.\ Rev.\ D {\bf 89}, no. 10, 103523 (2014)
  [arXiv:1403.0438 [astro-ph.CO]].


\bibitem{Hwang:2009zj}
  C.~G.~Park, J.~c.~Hwang, J.~h.~Lee and H.~Noh,
  Phys.\ Rev.\ Lett.\  {\bf 103}, 151303 (2009)
  [arXiv:0904.4007 [astro-ph.CO]].


\bibitem{Noh:2004bc}
  H.~Noh and J.~c.~Hwang,
  Phys.\ Rev.\ D {\bf 69}, 104011 (2004)
  [astro-ph/0305123].


\bibitem{Hwang:2007ni}
  J.~c.~Hwang and H.~Noh,
  Phys.\ Rev.\ D {\bf 76}, 103527 (2007)
  [arXiv:0704.1927 [astro-ph]].


\bibitem{Hwang:2005he}
  J.~c.~Hwang and H.~Noh,
  Phys.\ Rev.\ D {\bf 72}, 044012 (2005)
  [gr-qc/0412129].


\bibitem{eos}
  M.~Chevallier and D.~Polarski,
  Int.\ J.\ Mod.\ Phys.\ D {\bf 10}, 213 (2001)
  [gr-qc/0009008]~;
  E.~V.~Linder,
  Phys.\ Rev.\ Lett.\  {\bf 90}, 091301 (2003)
  [astro-ph/0208512].


\bibitem{Lewis:1999bs}
  A.~Lewis, A.~Challinor and A.~Lasenby,
  Astrophys.\ J.\  {\bf 538}, 473 (2000)
  [astro-ph/9911177].


\bibitem{Jeong:2006xd}
  D.~Jeong and E.~Komatsu,
  Astrophys.\ J.\  {\bf 651}, 619 (2006)
  [astro-ph/0604075].


\bibitem{g-inv}
  J.~Yoo, A.~L.~Fitzpatrick and M.~Zaldarriaga,
  Phys.\ Rev.\ D {\bf 80}, 083514 (2009)
  [arXiv:0907.0707 [astro-ph.CO]]~;
  J.~Yoo,
  Phys.\ Rev.\ D {\bf 82}, 083508 (2010)
  [arXiv:1009.3021 [astro-ph.CO]].


\bibitem{obsPg}
  C.~Bonvin and R.~Durrer,
  Phys.\ Rev.\ D {\bf 84}, 063505 (2011)
  [arXiv:1105.5280 [astro-ph.CO]]~;
  A.~Challinor and A.~Lewis,
  Phys.\ Rev.\ D {\bf 84}, 043516 (2011)
  [arXiv:1105.5292 [astro-ph.CO]].


\bibitem{delta-redshift}
  N.~Kaiser,
  Mon.\ Not.\ Roy.\ Astron.\ Soc.\  {\bf 227}, 1 (1987)~;
  A.~F.~Heavens, S.~Matarrese and L.~Verde,
  Mon.\ Not.\ Roy.\ Astron.\ Soc.\  {\bf 301}, 797 (1998)
  [astro-ph/9808016].


\bibitem{Kaiser:1984sw}
  N.~Kaiser,
  Astrophys.\ J.\  {\bf 284}, L9 (1984).


\bibitem{proper time}
  D.~Jeong, F.~Schmidt and C.~M.~Hirata,
  Phys.\ Rev.\ D {\bf 85}, 023504 (2012)
  [arXiv:1107.5427 [astro-ph.CO]]~;
  J.~Yoo,
  Phys.\ Rev.\ D90 (2014) 12, 123507 arXiv:1408.5137 [astro-ph.CO].


\bibitem{prospect}
  D.~H.~Weinberg, M.~J.~Mortonson, D.~J.~Eisenstein, C.~Hirata, A.~G.~Riess and E.~Rozo,
  Phys.\ Rept.\  {\bf 530}, 87 (2013)
  [arXiv:1201.2434 [astro-ph.CO]]~;
  L.~Amendola {\it et al.}  [Euclid Theory Working Group Collaboration],
  Living Rev.\ Rel.\  {\bf 16}, 6 (2013)
  [arXiv:1206.1225 [astro-ph.CO]].



\end{thebibliography}
\end{document}